\documentclass[aps,prl,twocolumn,showpacs,superscriptaddress,groupedaddress]{revtex4}  
\usepackage{graphicx}  
\usepackage{dcolumn}   
\usepackage{bm}        
\usepackage{amssymb}   

\usepackage{color}

\newcommand{\imp}{\mathcal{I}}

\begin{document}

\title{Anomalous impact in reaction-diffusion models}
\author{I.~Mastromatteo} \affiliation{Centre de Math\'ematiques Appliqu\'ees, CNRS, UMR7641, Ecole Polytechnique, 91128 Palaiseau, France.}
\author{B.~T\'oth} \author{J.-P.~Bouchaud} \affiliation{Capital Fund Management, 23-25 Rue de l'Universit\'e, 75007 Paris, France.}
\date{\today}

\begin{abstract}
We generalize the reaction-diffusion model $A + B \to \emptyset$ in order to study the impact of an excess of $A$ (or $B$) at the reaction front. 
We provide an exact solution of the model, which shows that linear response breaks down: the average displacement of the reaction front 
grows as the {\it square-root} of the imbalance. We argue that this model provides a highly simplified but generic framework to understand the square-root 
impact of large orders in financial markets. 
\end{abstract}

\maketitle

In most systems, small perturbations induce proportionally small responses: this is the {\it linear response} regime. Critical systems are exceptions to this general rule: long-range correlations
make these systems particularly fragile. A well known example of anomalously large response is the magnetic susceptibility close to the para/ferromagnetic transition. In fact, exactly at the transition, 
the magnetisation $M$ is zero in the absence of an external force (the magnetic field $H$), but behaves when $H \to 0$ as $M \sim H^{\delta}$ with $\delta < 1$ (for example $\delta=1/3$ in mean-field, 
see e.g.~\cite{Goldenfeld:1992}). 
The fact that $\delta < 1$ is tantamount to saying that the linear response coefficient $\lim_{H \to 0} M/H$ diverges at criticality, indicating anomalous fragility. Conversely, the observation 
of a diverging linear response suggests a non-trivial underlying organisation of the system. This is partly the reason why the recently reported universal anomalous impact of small trades in financial markets has triggered a spree of activity (see e.g. \cite{Almgren:2005,Farmer:2011,Toth:2011,Iacopo:2013} and refs.\ therein). Market impact is not only a problem of paramount importance for finance practitioners (for whom market impact amounts to trading costs), 
it also relates to one of the most fundamental questions in theoretical economics: \emph{why and how do prices change}? 
Market impact is at the core of the sophisticated mechanism through which markets absorb trading information as an input and produce prices as an output \cite{Bouchaud:2008,Bouchaud:2010}. 
The failure of such a mechanism can have dramatic consequences for society, ranging from market inefficiencies to full-fledged crashes (see e.g.~\cite{Bouchaud:2011} and refs.\ therein).

More precisely, by ``market impact'' we mean the average price change $\imp$ after the sequential execution of a total volume $Q$ of contracts  (which we call \emph{meta-order}).
Contrary to the models customarily employed in the field of theoretical economics \cite{Kyle:1985}, in which $\imp$ is traditionally assumed to be a linear function of $Q$, a 
growing consensus in the empirical literature indicates that impact follows a concave law, that is well-described by the so called ``square-root'' impact formula:
\begin{equation}
\label{eq:sqrt_fin}
	\imp = Y \sigma_D \left( \frac{Q}{V_D} \right)^\delta \; ,
\end{equation}
where $\delta$ is an exponent in the range $0.4-0.7$, $\sigma_D$ and $V_D$ are the respectively the daily price fluctuations and the daily traded volume \cite{Barra:1997,Almgren:2005,Moro:2009,Toth:2011,Iacopo:2013}. $Y$ is a dimensionless coefficient which is found to be of order 1.

As mentioned above, the fact that $\delta < 1$ indicates that markets are inherently fragile: vanishingly small traded volumes are expected to have a disproportionate impact on prices. Even more surprisingly, the law appears to be \emph{universal}, as it is to a large degree independent of details such as the type of contract traded, the geographical position of the market venue, the period of time 
in which trading takes place or the strategy used to execute the order \cite{Toth:2011}
(by universality we do not just mean the form of Eq.~(\ref{eq:sqrt_fin}) but also the value of the dimensionless coefficient $Y$).
It appears to be extremely robust against microstructural changes; for example the rise of high-frequency trading (HFT) in the last ten years seems to have had very little effect on the validity of Eq.~(\ref{eq:sqrt_fin}). Such a universality is the main reason why one should expect simple models to be able to reproduce the square-root law. If the relevant properties of the market are included in a stylized model, the low-frequency properties of the dynamics (say, from some hours to a few days) should be correct even if the high-frequency (say, below one minute) description is inaccurate or not realistic.

In this spirit, we propose here a coarse-grained model of the market much inspired by \cite{Bak:1997,Tang:1999}, which relies on two fundamental ingredients in order to describe market dynamics: 
\emph{i)}~participants place and update orders to buy (sell) at prices as low (high) as possible; \emph{ii)}~market clearing (buy orders and sell orders annihilate each other when at the same price). Therefore, we postulate, as in \cite{Toth:2011}, the existence of a \emph{latent order book} (modeled as a one-dimensional grid of length $L$) encoding the trading intentions of the market participants: in this setting each price level $x$ can be populated by particles of two types ($B$ and $A$), representing respectively the intended orders to buy (\emph{bids}) and to sell (\emph{ask}). In practice, such book can be seen as a proxy for the supply and demand curves at the intra-day scale. Such point of view tacitly assumes \emph{segregation} among the two particle types (i.e., supply and demand curves do not overlap), and implicitly enforces the presence of a finite \emph{spread} separating the highest bid (the rightmost $B$ particle) and the lowest ask (the leftmost $A$ particle) through a market clearing condition.

The stochastic dynamics that we propose for the particles populating the book consists of a hopping process for both type of particles (each particle can jump either right or left with probability $D$ per unit time) and of a reaction process mimicking the market clearing condition: particles at the same site will have a probability $\lambda$ per unit time to start a reaction process (we will eventually consider the limit $\lambda \to + \infty$). The reaction process may have three different outcomes, chosen at random according to the value of two parameters $p$ and $m$:
\begin{eqnarray}
\label{eq:annih}
A+B \to \emptyset \quad &\textrm{w. prob.}& \quad 1-p \\
A+B \to B \quad &\textrm{w. prob.}& \quad p \; \frac{1+m}{2} \\
A+B \to A \quad &\textrm{w. prob.}& \quad p \; \frac{1-m}{2}  \; .
\end{eqnarray}
For $p=0$, this boils down to the model studied in \cite{Bak:1997,Tang:1999}, but this setting is too restrictive as it does not allow one to introduce a {\it bias} $m$, which is of course a crucial ingredient to study impact.
In fact, the events associated with $p>0$ can be interpreted as due to the action of an additional agent, who adds to the system an extra bid particle (with probability $(1+m)/2$) or an extra ask particle (with probability $(1-m)/2$). The lack of a conservation law for the difference between the number of buy and sell particles is then explained by the imbalance introduced by such extra agent.
Finally, we suppose that a flux of particles per unit time $J_B=J_A=J$ (of type $B$ and $A$) are inserted at the boundaries (respectively at sites $1$ and $L$). Hence, the system lies in a non-equilibrium state due to the presence of an external particle pressure, representing the flux of orders coming from new participants, that can become interested in entering the market. The model will only make sense if the 
results do not depend on $L$, which is to a large extent arbitrary.

The model described above leads in continuous approximation to the following dynamics:
\begin{eqnarray}
\label{eq:dynamics}
\frac{\partial \langle b(x,t) \rangle}{\partial t} &=& D \frac{\partial^2 \langle b(x,t) \rangle}{\partial x^2} - \lambda u_A \langle a(x,t) b(x,t) \rangle \\
\frac{\partial \langle a(x,t) \rangle}{\partial t} &=& D \frac{\partial^2 \langle a(x,t) \rangle}{\partial x^2} - \lambda u_B \langle a(x,t) b(x,t) \rangle \; ,
\end{eqnarray}
where $a(x,t)$ and $b(x,t)$ are the densities of particles of type $A$ and $B$, and $u_A = 1-p(\frac{1+m}{2})$ and $u_B = 1-p(\frac{1-m}{2})$. In this limit the conditions at the boundary become the Neumann boundary conditions
\begin{eqnarray}
J = - D \frac{\partial \langle b(x,t) \rangle }{\partial x} \bigg|_{x=0} &\quad & \phantom{-}0\; = - D \frac{\partial \langle b(x,t) \rangle }{\partial x} \bigg|_{x=L} \\
0 = - D \frac{\partial \langle a(x,t) \rangle }{\partial x} \bigg|_{x=0} &\quad & -J = - D \frac{\partial \langle a(x,t) \rangle }{\partial x} \bigg|_{x=L}
\end{eqnarray}
This model is extremely hard to solve in one dimension due to the presence of strong correlations among the particle positions \cite{Ben-Naim:1992,Cornell:1993}. Whereas in higher dimension (or in the small coupling regime $\lambda J^{-1/2} D^{-1/2} \ll 1$) the mean field approximation $\langle a b \rangle = \langle a \rangle\langle b \rangle$ is quite accurate, in one dimension and in the large coupling regime $\lambda J^{-1/2} D^{-1/2} \gg 1$ (which is relevant here), interactions are too strong for the mean-field prediction to be even qualitatively correct \cite{Ben-Naim:1992,Cornell:1993}. In that case, even in the simpler case $p=0$, it is necessary to rely on approximate results obtained by using sophisticated renormalization group techniques \cite{Cardy:1996} or to resort to numerical simulations \cite{Araujo:1993,Cornell:1995}.

In our setting, the symmetric case $p=0$ corresponds to the case in which the flux of the market is balanced, i.e., no meta-order is being executed. Hence, it represents the market unperturbed state, and it is then worth to underline its main features. First, we remark that in the symmetric case $u_A=u_B$, due to the conservation law for the difference of $A$ and $B$ particles, the combination $\varphi = b - a$ follows a diffusion equation of the type $\partial_t \varphi = D \partial_{xx}^2 \varphi$, subject to the boundary condition $-D \partial_x \varphi |_{x=0,L} = J$. The stationary state is immediate to compute and results in a linear density profile:
\begin{equation}
\label{eq:stat_phi}
\varphi_{st}(x) = -(J/D)(x-L/2).
\end{equation}
Second, the interface of the model $x^{*}_t$ (corresponding to the traded price) diffuses anomalously: while at large times the boundaries obviously confine the system between $x=0$ and $x=L$, in the small time regime $tD/L^2 \ll 1$ the interface diffuses very slowly, as the law of $|x^{*}_t-x^{*}_0|$ is found to be compatible with $\sim \log t$ (as opposed to the case $J=0$ considered in~\cite{Cardy:1996} which leads to $|x^{*}_t-x^{*}_0|\sim t^{1/4}$). In particular for $L\to\infty$ the interface -- and hence the mid-price -- is sub-diffusive.
Despite being at odds with empirical observations of actual financial markets, sub-diffusion of the price within the model is expected from the confining effect of the order book itself: the diffusive nature of prices in a financial market (namely, the fact that for times larger than a few trades one has $|x^{*}_t-x^{*}_0|\sim t^{1/2}$) is enforced by  \emph{strategic} interactions, a mechanism which we have chosen not to include in the present version of our model (see \cite{Toth:2011,Iacopo:2013} for a detailed discussion of this point). 

\begin{figure}
 \centering
   \includegraphics{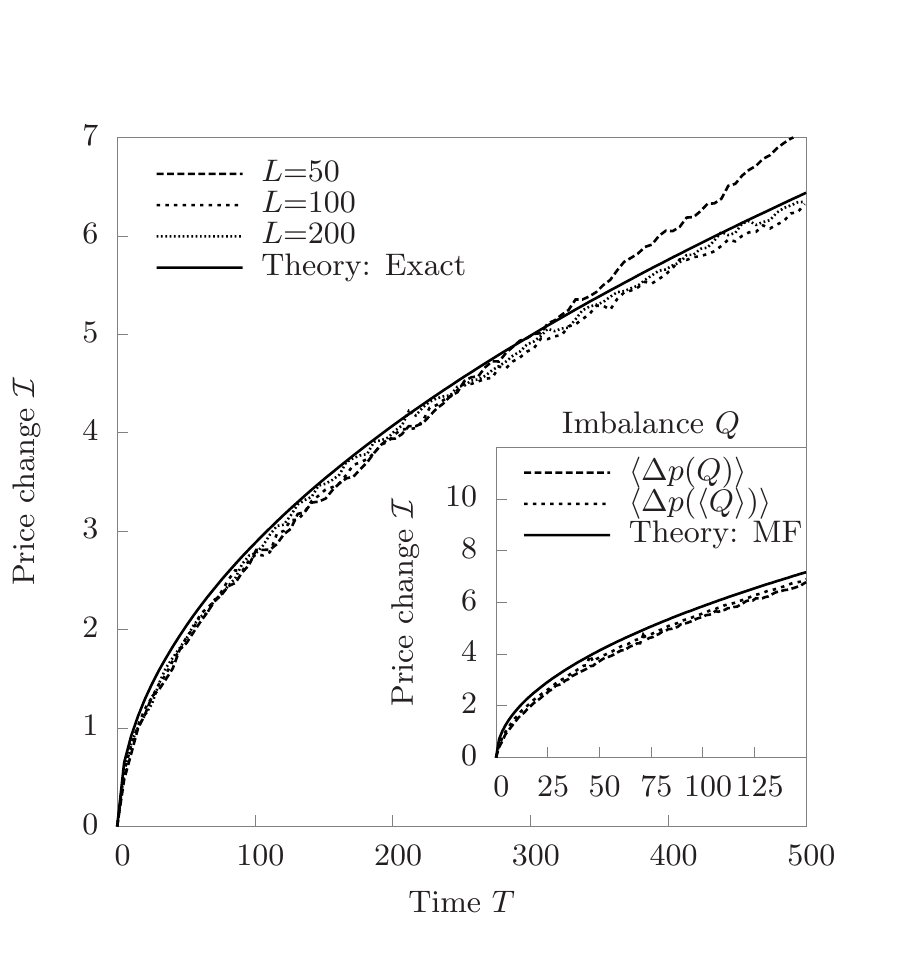}
 \caption{
  (\emph{Main figure}) Average change in the position of the mid-point $\imp = \langle x^{*}_t \rangle - L/2 $ after a perturbation of duration $T$. We compare the results of simulations of systems of different length (dashed lines) with the analytical prediction valid in the limit $L\to\infty$ (solid line) finding very good agreement. We have used the parameters $J=D=1$, $p=0.5$ and $m=0.75$. The limit $\lambda\to\infty$ is enforced by setting $\lambda=10^3$. Also notice the crossover of the curve to the linear regime (indicating $Dt/L^2 \gtrsim 1$) appearing in the curve for $L=50$.
  (\emph{Inset}) Average change in the mid-point position $\imp$ compared against the volume imbalance $Q$ for a simulated system of length $L=100$ (dashed lines). The solid line indicates the mean-field (MF) estimate predicted by Eq.~(\ref{eq:avg_imb}). We have chosen the parameters $D=J=1$, $\lambda=1000$, $p=1$ and $m=0.5$.
  }
  \label{fig:impact}
\end{figure}

The goal of the present discussion is to investigate the change in the interface position due to an imbalance in the order flux, i.e.~the case $p \neq 0$, $m \neq 0$. We model such imbalance by supposing that the system, after being prepared in the symmetric stationary state at time $t=0$, is subject to a sudden change of the values $p$ and/or $m$ controlling the imbalance parameters $u_A,u_B$ until a time $t=T$. In that case, it is convenient to study the evolution of the linear combination $\psi = u_B b - u_A a$, which again follows a simple diffusion equation:
\begin{equation}
\label{eq:diff_psi}
\frac{\partial \langle \psi (x,t)\rangle}{\partial t} = D \frac{\partial^2 \langle \psi (x,t)\rangle}{\partial x^2} \; , 
\end{equation}
with boundary conditions
\begin{equation}
\label{eq:bound_psi}
J u_B= - D \frac{\partial \langle \psi(x,t) \rangle }{\partial x} \bigg|_{x=0} \quad  J u_A = - D \frac{\partial \langle \psi(x,t) \rangle }{\partial x} \bigg|_{x=L}  \;.
\end{equation}
The interest in the field $\psi$ lies in the fact that for $\lambda \to \infty$ its zeroes {\it coincide} with the zeroes of the field $\varphi = b - a$. Hence, by identifying the average price change $\langle x_t^{*} \rangle$ with the point verifying $\langle \varphi( \langle x_t^{*} \rangle ,t) \rangle = 0$, it is possible to connect the solution of Eq.~(\ref{eq:diff_psi}) with the expected position of the interface at a time $t=T$ after the initial perturbation.
The solution of Eq.~(\ref{eq:diff_psi}) subject to the boundary conditions~(\ref{eq:bound_psi}) and the initial conditions~(\ref{eq:stat_phi}) is:
\begin{eqnarray}
  \label{eq:sol_psi}
  f(y,\tau) &=&  \frac{1}{12} (u_B-u_A) - \frac{u_B + u_A}{2} y + \frac{u_B-u_A}{2}y^2 \\
  &+& \nonumber
  (u_B-u_A) \tau  -  \frac{u_B-u_A}{2} \sum_{n=1}^\infty \frac{\cos(2\pi n y)}{\pi^2 n^2} e^{- 4 \pi^2 n^2 \tau}  \; ,
\end{eqnarray}
where we have defined the dimensionless variables
\begin{eqnarray}
  \label{eq:adim}
  \tau &=& D T / L^2 \\
  y    &=& x/L - 1/2 \\
  f(y,\tau) &=& \frac{D}{JL} \psi(y(x),T(\tau)) \; .
\end{eqnarray}
An inspection of Eq.~(\ref{eq:sol_psi}) at $\tau=0$ reveals that the motion of the interface is due to the discontinuous shape of $\psi(x,0)$ right after the perturbation: the smooth stationary shape of $\varphi_{st}(x)$ is mapped into the piecewise linear function  $\psi(x,0)$ . Additionally, the boundary conditions for $\psi$ are asymmetric, implying that in the modified coordinates the side pushing the interface with more pressure encounters a milder resistance on the other side in terms of particle density.
The trajectory of the average mid-point $\langle x^{*}_T \rangle = L(1/2+y_\tau^{*})$ can be computed by exploiting the relation
\begin{equation}
  \label{eq:mid_point_rel}
  0=\frac{d}{d\tau} f(y^{*}_\tau,\tau) = \frac{\partial f}{\partial y} \dot y^{*}_\tau + \frac{\partial f}{\partial \tau} \; ,
\end{equation} 
while the partial derivatives can be extracted from Eq.~(\ref{eq:sol_psi}), which implies:
\begin{eqnarray}
  \label{eq:deriv_sol}
    \frac{\partial f}{\partial \tau} &=& (u_B-u_A) \Theta_3(\pi y,e^{-4\pi^2\tau} ) \\
  \frac{\partial f}{\partial y}    &=& - \frac{u_B + u_A}{2} + (u_B-u_A) \int_0^y dy^\prime \Theta_3(\pi y^\prime, e^{-4\pi^2\tau} ) \; .  \nonumber
  \end{eqnarray}
where $\Theta_3(z,q)$ is the Jacobi theta function of the third kind. The above expressions can be used to solve Eq.~(\ref{eq:mid_point_rel}) with respect to $\dot y^{*}_\tau$.
A small $\tau$ expansion for $\Theta_3(\pi y,e^{-4\pi^2 \tau} ) $ leads finally to a differential equation for the trajectory $y_\tau^{*}$, whose solution is
\begin{equation}
  \label{eq:mid_point}
 y_\tau^{*} =  2 \alpha\left( u_B/u_A \right) \tau^{1/2} \; ,
\end{equation}
where the function $\alpha(z)$ satisfies the transcendental equation:
\begin{equation}
  \label{eq:alpha}
  \alpha(z) \left( \frac{z+1}{z-1} - \textrm{erf}[\alpha(z)] \right) - \frac{1}{\sqrt{\pi}} e^{-\alpha^2(z)}.
\end{equation}
Eqs.~(\ref{eq:mid_point}) and~(\ref{eq:alpha}) are our central result: they state that the average change in the interface position grows as the square root of the rescaled time. Moreover, when putting back the original units, one finds that $\langle x^{*}_T \rangle - L/2 = 2 \alpha (DT)^{1/2}$, {\it independent} of $L$. This means that in the infinite size limit $DT/L^2 \to \infty$ the impact is unaffected by the long size behavior of the system. Finally, in this regime Eq.~(\ref{eq:mid_point}) becomes exact, as the large $L$ regime corresponds to the small $\tau$ limit. Numerical simulations of the model have been performed in this regime, finding perfect agreement with Eq.~(\ref{eq:mid_point}) (see Fig.~\ref{fig:impact}).

In order to relate this findings to empirical results on market impact Eq.~(\ref{eq:sqrt_fin}), we need to link the variation of the mid-price $\langle x^{*}_T \rangle$ to the executed volume $Q$. According to the financial interpretation suggested above, $p>0$ represents the action of an additional agent which for $m\neq0$ is introducing a bias in the volume imbalance.
Hence it is natural to identify such bias as the volume $Q$ executed by the agent.
Its average is equal to $\langle  Q\rangle= \int dx \, \left<(b - a) \right> = D\int dt \,(\langle\partial_x a \rangle_{x=x^{*,+}} + \langle \partial_x b\rangle_{x=x^{*,-}})$, the average number of $A$ particles that reached the interface minus the number of $B$ particles that touched the reaction zone.
Another quantity of interest is $\langle V\rangle= D\int dt \,(\langle\partial_x a \rangle_{x=x^{*,+}} - \langle \partial_x b\rangle_{x=x^{*,-}})$, which is equal to the total number of particles that reacted. An accurate approximation of $\langle Q \rangle$ and $\langle V \rangle$ can be obtained by mapping Eq.~(\ref{eq:deriv_sol}) on the original coordinate system, so to integrate in time the fluxes through the interface. Exploiting again the properties of the Jacobi theta function of the third kind, one finds that
\begin{eqnarray}
\label{eq:avg_imb}
\langle Q \rangle &=& \beta(u_B /u_A) (J T) \\
\label{eq:avg_exec}
\langle V \rangle &=& \gamma(u_B /u_A) (J T) \; ,
\end{eqnarray}
where the functions $\beta(z)$ and $\gamma(z)$ are given by
\begin{eqnarray}
\label{eq:beta}
\beta(z) &=& \frac 1 {2z} \left[ (z^2-1) -  \textrm{erf}[\alpha(z)] (z-1)^2 \right] \\
\label{eq:gamma}
\gamma(z) &=& \frac 1 {2z} \left[ (z+1)^2 -  \textrm{erf}[\alpha(z)] (z^2-1) \right] \; .
\end{eqnarray}
Eq.~(\ref{eq:avg_imb}) leads to an approximate estimate of the impact of the type $\imp = 2 \alpha (Q D/\beta J)^{1/2}$, which is in very good agreement with the simulation results shown in the inset of Fig.~\ref{fig:impact}.
Eq.~(\ref{eq:gamma}) can be used to characterize the imbalance parameter $z=u_B/u_A$ as a function of the \emph{participation rate} of the additional agent $\phi = 2Q / (Q + V) $, whose average is equal in mean-field approximation to
\begin{equation}
\label{eq:part_rate}
\langle \phi  (z) \rangle= \frac{2\beta(z) }{\beta(z) + \gamma(z)} \; .
\end{equation}
Eqs.~(\ref{eq:beta}) and~(\ref{eq:gamma}) can also be used to associate the $Y$ term appearing in Eq.~(\ref{eq:sqrt_fin}) with the combination $Y(z) = \alpha(z) \beta^{-1/2}(z)$. For small $\phi$ this is approximately equal to $Y \approx (\phi/4\pi)^{1/2}$, at odds with empirical observations.

\begin{figure}
  \centering
 \includegraphics{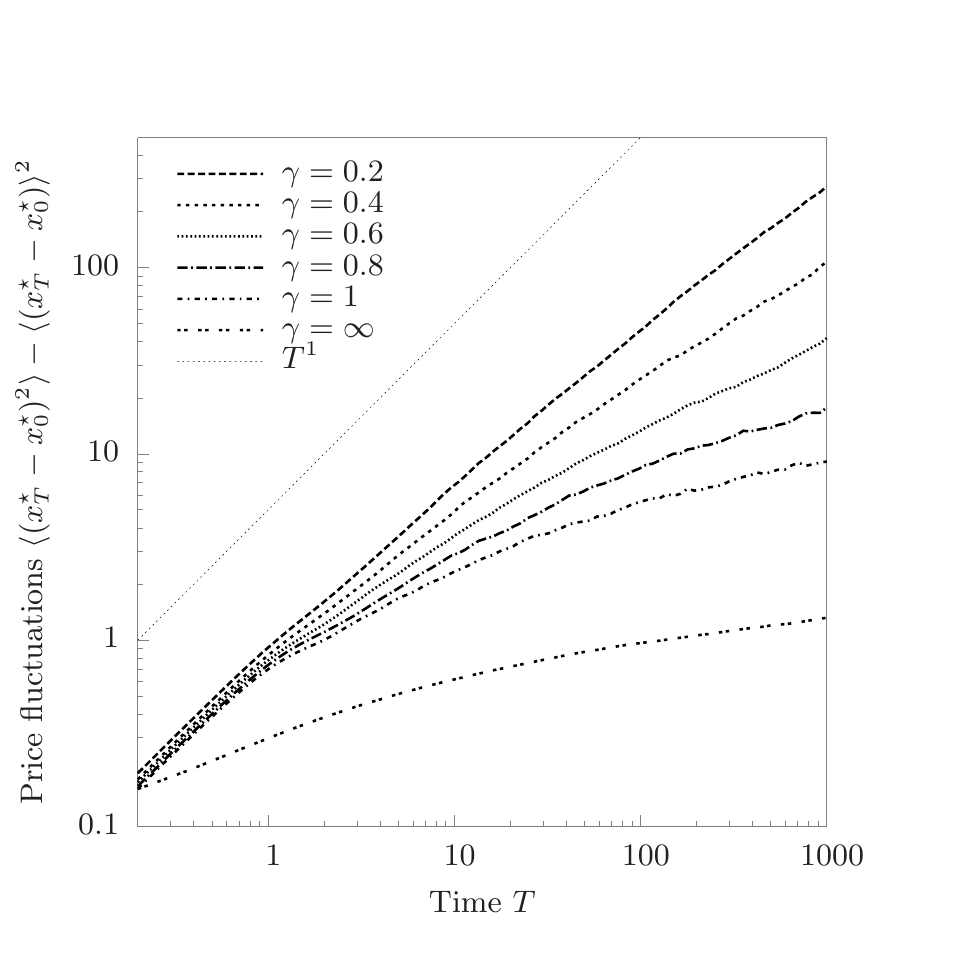}
  \caption{
Fluctuations in the interface position for a modified model in which the terms $u_A$ and $u_B$ are random variables.
In particular we change Eq.~(\ref{eq:annih}) by choosing with probability $1-p$ the sign of the reaction ($A+B\to $ either $A$ or $ B$) according to a zero-mean, long range correlated process with tail exponent $\gamma$.
We find that the diffusion properties of the model change even though the impact properties are unaffected.
We plot the variance of the interface position for different values of $\gamma$ for the set of parameters $L=400$, $J=D=1$, $\lambda=1000$ and $p=m=0$.
}
  \label{fig:diffusion}
\end{figure}

All the above results hold in an extremely broader context: \emph{(i)}~if drifts term of the type $\mu \langle \partial_x a \rangle,\mu \langle \partial_x b \rangle$, or if decay terms $-\nu \langle a\rangle,-\nu\langle b\rangle$ are added to Eq.~(\ref{eq:dynamics}), then an extra timescale will implicitly be induced in the model. In this case Eqs.~(\ref{eq:mid_point}),~(\ref{eq:beta}) and~(\ref{eq:gamma}) will still provide a correct description of the system in regime of small times. Secondly, \emph{(ii)} when changing the reaction term $\lambda  u_{A/B} \,a\, b$ to any other symmetric combination of $a$ and $b$, the equation for $\psi$ will be unaltered. This implies that by appropriately tuning the reaction term, it is possible to change the diffusion properties of the system 
all the way from $\log t$ to $t^{1/2}$ {\it without affecting the square-root impact law}, Eq.~(\ref{eq:mid_point}) (see Fig.~\ref{fig:diffusion}).

In this paper, we have provided an analytically tractable implementation of the type of system proposed in \cite{Toth:2011}: in our model market clearing indeed induces a locally linear (V-shaped) 
liquidity profile close to the traded price, which in turn induces a square root impact shape, as suggested by the mean-field argument in \cite{Toth:2011}. However, it is highly non-trivial that 
such a mean-field argument gives the correct answer since the fluctuations in the interface position are in fact found to be much larger than the impact itself. It is therefore quite important to 
have a model where the ``square-root'' impact can be established analytically (rather than numerically, as in \cite{Toth:2011,Iacopo:2013}). Even though the exact predictions of our stylized model might depend on the actual choice of the reaction parameters, our results suggest that in a one-dimensional system of annihilating particles, a concave dependence of the interface position on the flux imbalance should be regarded as the rule, rather than as the exception. This confirms that very generic features (diffusion and market clearing condition) are, as surmised in \cite{Toth:2011}, sufficient to 
explain the anomalous reaction of prices to volume imbalances. As emphasized in \cite{Toth:2011} and recalled in the introduction, this also means that markets are ``critical'', i.e.\ generically close to an instability since the liquidity is vanishingly small in the vicinity of the current price. Liquidity {\it fluctuations} are thus bound to play a crucial role, and we expect these fluctuations
to be at the heart of the turbulent dynamics of financial markets \cite{Lillo:2005, Bouchaud:2011, Toth:2011}.

{\it Acknowledgements} We thank R. Benichou, J.de Lataillade, C. Deremble, J. Donier, D. Farmer, J. Gatheral, J. Kockelkoren, P. Kyle, Y. Lemp\'eri\`ere, F. Lillo, M. Potters and H. Waelbroeck for many discussions on these issues. 
This research benefited from the support of the ``Chair Markets in Transition'', under the aegis of ``Louis Bachelier Finance and Sustainable Growth'' laboratory, a joint initiative of \'Ecole Polytechnique, Universit\'e d'\'Evry Val d'Essonne and F\'ed\'eration Bancaire Fran\c{c}aise.

\end{document}